\documentclass[sigconf, nonacm]{acmart}

\usepackage{caption}
\usepackage{tikz}
\usepackage{filecontents}
\usepackage{balance}
\usepackage{multicol}
\usepackage{multirow}
\usepackage{booktabs}
\usepackage{subfigure}
\usepackage{booktabs}
\usepackage{enumitem}
\newcommand{\header}[1]{\vspace{1mm}\noindent\textbf{#1.}}

\usepackage{color}
\usepackage{xcolor}
\usepackage{colortbl,booktabs}

\newcommand{\LQoCo}{\texttt{L-QoCo}\xspace}
\newcommand{\NoCoTo}{\texttt{No-CoTo}\xspace}
\newcommand{\CoTo}{\texttt{CoTo}\xspace}
\newcommand{\SB}{\texttt{L-QoCo-AB}\xspace}
\newcommand{\DK}{\texttt{L-QoCo-DK}\xspace}
\newtheorem{definition}{Definition}

\newcommand{\eat}[1]{}
\usepackage{makecell}
\usepackage[utf8]{inputenc}
\usepackage{amsmath}
\usepackage[linesnumbered,ruled,vlined]{algorithm2e}
\usepackage{algorithmic}

\SetKwInput{KwInput}{Input}                % Set the Input
\SetKwInput{KwOutput}{Output}              % set the Output

\begin{document}
\title{\LQoCo: Learning to Optimize Cache Capacity Overloading \\in Storage Systems}

%%
%% The "author" command and its associated commands are used to define the authors and their affiliations.

% \author{\IEEEauthorblockN{
% Ji Zhang\IEEEauthorrefmark{1},
% Xiyao Zhou\IEEEauthorrefmark{1},
% Xijun Li\IEEEauthorrefmark{2}\IEEEauthorrefmark{3},
% Mingxuan Yuan\IEEEauthorrefmark{2}, 
% Zhuo Cheng\IEEEauthorrefmark{1}, 
% Keji Huang\IEEEauthorrefmark{1}, 
% Lei Chen\IEEEauthorrefmark{4} 

% }
% \IEEEauthorblockA{
% \IEEEauthorrefmark{1}Huawei Technologies Co., Ltd\\
% \IEEEauthorrefmark{2}Huawei Noah's Ark Lab \\
% \IEEEauthorrefmark{3}MIRA Lab, University of Science and Technology of China\\
% \IEEEauthorrefmark{4}Hong Kong University of Science and Technology}
% }
\author{Ji Zhang$^{1}$, Xijun Li$^{2,}$$^{3}$$^{*}$, Xiyao Zhou$^{1}$, Mingxuan Yuan$^{1,}$$^{2}$, Zhuo Cheng$^{1}$, Keji Huang$^{1}$}
\affiliation{%
 \institution{$^{1}$Huawei Data Storage and Machine Vision Product Line, 
 $^{2}$Huawei Noah's Ark Lab,
 $^{3}$MIRA Lab, USTC
 }
}

% \author{\large{Ji Zhang, Xijun Li, Xiyao Zhou, , Mingxuan Yuan, Zhuo Cheng, Keji Huang, Yifan Li}}
% \affiliation{%
%       \institution{\large{Data Storage and Machine Vision Product Line}}
% }
% \affiliation{%
%   \institution{\large{Huawei Technologies Co., Ltd}}
% \affiliation{%
%   \institution{\large{MIRA Lab, USTC}}
\begin{abstract}

%It is crucial for storage systems to maintain good and stable performance %(such as high throughput, low tail latency and throughput jitter)
%under varieties of workloads. However, maintaining such performance within the storage system is extremely challenging. This is because the real-world workloads might be very time-varying. Besides, the architecture of the current storage system is so complex and coupled that engineers are unable to directly find the bottleneck and tune it. Empirically, we find that the optimization of cache capacity overload might be a concise and effective solution to the performance enhancement of the whole storage system. %Nevertheless, few previous works about performance optimization of storage systems pay attention to this problem.

\footnote{Xijun Li is the corresponding author. This paper has been accepted by DAC 2022 and will appear in the proceeding of DAC 2022.}Cache plays an important role to maintain high and stable performance (i.e. high throughput, low tail latency and throughput jitter) in storage systems. Existing rule-based cache management methods, coupled with engineers' manual configurations, cannot meet ever-growing requirements of both time-varying workloads and complex storage systems, leading to frequent cache overloading. 

In this paper, we for the first time propose a light-weight learning-based cache bandwidth control technique, called \LQoCo which can adaptively control the cache bandwidth so as to effectively prevent cache overloading in storage systems.  Extensive experiments with various workloads on real systems show that %~\footnote{Huawei storage is one of the world's leading storage vendor which has provided professional storage services to over 12,000 enterprise-class customers in global.}  
\LQoCo, with its strong adaptability and fast learning ability, can adapt to various workloads to effectively control cache bandwidth, thereby significantly improving the storage performance  (e.g. increasing the throughput by 10\%-20\% and reducing the throughput jitter and tail latency by 2X-6X and 1.5X-4X, respectively, compared with two representative rule-based methods).

%In this work, we propose a light-weight learning-based control framework for cache capacity overload problems in storage systems, called \LQoCo. To the best of our knowledge, it is the first work to solve the  problem using reinforcement learning. Inside \LQoCo, an \textit{adaptive bound strategy} is elaborately designed to further improve the robustness and learning efficiency of the framework. To verify the effectiveness of \LQoCo on performance enhancement, extensive experiments in the native storage system on Linux and the Huawei storage product OceanStor%~\footnote{Huawei storage is one of the world's leading storage vendor which has provided professional storage services to over 12,000 enterprise-class customers in global.} 
%under different workloads were performed. Compared with two rule-based practical overload control mechanisms, the results reveal that \LQoCo posses strong adaptability and fast learning ability, which significantly improves the performance of the storage system (throughput $10\%\sim20\%\uparrow$, throughput jitter $2X\sim6X \downarrow$, and tail latency $1.5X \sim 4X \downarrow$).
\end{abstract}

\maketitle

\pagestyle{plain}
%\input{secs/abstract}
% !TEX root = main.tex
%\vspace{-1.2em}
% \begin{figure}[h]
% \centering
% \includegraphics[width=3.3in]{./fig/OverallStructure.pdf}
% \caption{The commonly-used centralized storage system architecture, comprising host-side servers, caching tier and storage backend.}
% \label{OverallStructure}
% \end{figure}

\vspace{-0.6em}
\section{Introduction} \label{intro}
%%%1、先介绍集中式存储存储的架构，2、然后简要说明过载的问题（ext4上面和dorado上的不控的图摆出来），3、然后大致分析产生过载的因素，4、最后阐明我们要解决这个问题，5、贡献。6、文章组织
%With the explosive growth of information, modern datacenters are becoming larger and more complicated in order to store an ever-increasing amount and increasingly diverse type of data~\cite{wu2006design}. As a result, they are sensitive to tail latency and require high throughput~\cite{Inho2020, chen2020utree}. Figure~\ref{OverallStructure} shows the commonly used storage system architecture which consists of host-side system, networking tier, caching-tier and storage backend. It contains many components (e.g., workloads, networks, caches and disks) and algorithms (e.g., quality of service, log-structured store~\cite{o1996log}, cache prefetching, cache replacing, garbage collection and I/O scheduling) that make the storage systems become increasingly difficult to design, configure and manage. Therefore, maintaining sustainable high and stable performance especially under storage system overload is difficult.
% \xijun{testing}

With the explosive growth of data, storage systems are becoming larger and more complicated to store an ever-increasing amount and increasingly diverse type of data. %\textcolor{blue}{%In general, from the perspective of architecture, the modern storage system can be classified into two categories, centralized storage and distributed storage.} 
Storage systems are supposed to provide low tail latency with high and stable throughput~\cite{Shaohong2020} which users are very concerned about.  
% To meet the strict requirements of modern applications, they are sensitive to tail latency and require high and stable throughput~\cite{Inho2020, chen2020utree}.
%In this paper, the centralized storage architecture is mainly concerned. 
%The commonly-used storage system consists of host-side, caching-tier and storage backend\cite{Kan2021}. %such as the native storage system in Linux platform and our designed storage system by Huawei OceanStor.
% , with which our production-ready storage system is the same.  
% Note that our production-ready storage system is designed the same as this (See Figure~\ref{OverallStructure} in Appendix~\ref{productionsystem}). 
%In general, the host-side is comprised of general-purpose servers, which issues I/O requests to the caching-tier. Then these I/O requests are first cached in the caching-tier, followed by being persistently stored in the storage backend. 
Cache plays a vital role since it determines the overall performance of the storage system~\cite{Beckmann2017, Beckmann2016}. However, maintaining the stable and high performance of these storage systems poses great challenges to many rule-based cache management methods~\cite{FrancoJuliana2017, slimani2021service}.

% There are three basic cache writing policies: write-through, write-around and write-back (we described their difference in Figure~\ref{OverallStructure}). The commonly used cache writing policies in the storage systems are write-back, write-through and write-around (detailed difference see Figure~\ref{fig:poorperform}). In our paper we take write-back as an example for demonstrating. Note that our proposed framework also performs well in other two polices.  

\iffalse
\begin{figure*}[!t]\vspace{-.0em}
\centering
\subfigure[Without cache overload control in Linux. ]{
\includegraphics[width=3.0in]{./fig/linux_noCtrl.pdf}
}
\quad
\subfigure[Without cache overload control in Huawei OceanStor. ]{
\includegraphics[width=3.0in]{./fig/dorado_noCtrl.pdf}
}
\caption{The two systems without a cache overload control method show large throughput jitter and high tail latency.} 
\label{fig:nocontrol}
\end{figure*}
\fi

Cache capacity overloading (formal definition see DEFINITION~\ref{subsec:PDC}) is one of the common and major problems affecting the performance of cache in storage architectures. More specifically, the I/O requests are blocked at the host-side servers since the cache is about to be full, which might greatly degrade the performance (such as throughput and latency) of storage system.
%Recently, some researches of the cache access overload optimization are drawing increasing, where high concurrent and burst accesses overload the cache bandwidth requirements in hybrid storage architectures. 
%Although there is extensive research work on improving cache utilization and performance through cache admission policy~\cite{AdaptSize17,TinyLFU17,Flashield19,OTAE20,MLWP20} and cache replacement policy~\cite{LIRS02,ARC03,LARC16,CACHEUS21} optimization after cache capacity is exhausted, few researches focus on how to keep the cache from overloading. 
There are two practical commonly used ruled-based methods for optimizing this problem. The first exploits the bandwidth of capacity devices in hybrid storage architectures by bypassing the excess load to improve the overall performance. The representative practical case is the `bypass'~\footnote{https://bcache.evilpiepirate.org/BcacheGuide/} strategy. % on the native storage system in Linux platform 
%For the capacity of a cache is relatively large in especially enterprise storage systems, most of vendors 
Another uses heuristic methods (e.g., state machine) to limit the overall bandwidth at the entrance of the cache to optimize the problem. The representative practical case is the `\CoTo{}'~\footnote{https://support.huawei.com/enterprise/en/doc/EDOC1000084194?section=j01w\&topic\\Name=understanding-the-cache-performance} in Huawei storage product OceanStor. However, through some preliminary studies, we find that these rule-based methods for cache capacity overloading optimization still suffer from performance degradation (such as high throughput jitter and tail latency). %Through some preliminary studies, we infer that there is no feedback mechanism for these methods to perceive or learn from the change of storage system states and time-varying workloads, which makes them fail to perform well. More detailed preliminary observations and analysis are presented in Section~\ref{sec:PSM}. 

For further improving these practical rule-based methods, in this paper we propose the first online reinforcement learning (RL) based cache capacity overloading control framework called \LQoCo %The challenges of using RL for cache capacity overload optimization are as follows: \textbf{Firstly}, it is quite challenging to formulate the cache capacity overload problem as a \textit{Markov decision process}~\cite{Bellman34} (MDP), because RL is effective only when MDP assumption holds.
%\textbf{Secondly}, based on the MDP, how to design an efficient RL learning algorithm for cache capacity overload control because most of RL algorithms are sample-inefficient~\cite{pietquin2011sample}. \textbf{Lastly}, how to assure robustness of the control policy learned by the RL because the exploration~\cite{coggan2004exploration} of RL might lead to worse performance of storage systems. 
%In this paper, we addressed the above challenges 
by making the following contributions:

\begin{itemize}[leftmargin=*,noitemsep]
%\item We first formally define the cache capacity overload problem in the storage system. And then a rule-based cache capacity overload control mechanism called \CoTo is proposed to solve the problem within the practical storage system on Huawei OceanStor. \CoTo outperforms the system without any cache capacity overload control mechanisms for the cache. 
\item We formally define the cache capacity overloading in the storage system and formulate it as a MDP problem. Based on that, a learned capacity overloading control framework \LQoCo for cache is proposed. %To the best of our knowledge, it is the first work to solve the cache capacity overload problem utilizing online reinforcement learning (RL) techniques.
\item We design a sophisticated reward function and incorporate useful domain knowledge into \LQoCo which greatly accelerates the model learning of RL. Besides, an adaptive bound strategy is designed to further improve the robustness of \LQoCo. 

% to protect the storage system from violating a safe bandwidth range.
%\item We propose a safeguard boundary strategy in \LQoCo  , while allowing the system and the strategy to continuously adapt and learn, which is not affected by the unpredictable event caused by exploration operations in exploration \& exploitation dilemma of RL.

\item We implemented \LQoCo within two practical storage systems on different platforms. \LQoCo is compatible with both different storage architectures and hardware mediums as cache.
\item Extensive experiment results justify that \LQoCo is a light-weight capacity overloading control framework for the cache with strong adaptability under various workloads. It can significantly improve the performance of storage systems. %(throughput $10\%\sim20\%\uparrow$, throughput jitter $2X\sim6X \downarrow$, and tail latency $1.5X \sim 4X \downarrow$). 
%Our framework is open-sourced and publicly available on Github.~\footnote{The URL.}
\end{itemize}
%using the cache watermark level alone can not accurately evaluate the state of the whole storage system and lakes a feedback mechanism which
%The reminder of the paper is organized as follows: Section~\ref{sec:PSM} presents our preliminary study (contribution \#1) and motivation. The detailed design and architecture of our RL-based \LQoCo (contribution \#2, \#3, and \#4) are discusses in Section~\ref{sec:PS}. The experimental results (contribution \#5) are explained in Section~\ref{sec:ES}. Section~\ref{sec:RW} provides substantial related work, while we close the paper with the conclusions and future works in Section~\ref{sec:C}.

% !TEX root = main.tex

\vspace{-0.6em}
\section{Background and Motivation} \label{sec:PSM}

%In this section, the background is formally presented. Then the preliminary results and motivation are described at length. 

% In this section, we first give the problem definition for cache overload in a storage system, then conduct two preliminary experiments in Linux and our production-ready platform to demonstrate the necessity of cache overload control.  Finally, we introduce our motivation to design a learned overload control framework for cache using RL. 

\iffalse
\begin{figure}\vspace{-2.em}
\centering
%\subfigure[Duration of cache being full with/without capacity overload method in Linux and OceanStor.]
{
\includegraphics[width=3.1in]{./fig/cache_full_stat_new.pdf}
}
%\quad
%\subfigure[Duration of cache being full with/without capacity overload method in OceanStor. ]{
%\includegraphics[width=3.1in]{./fig/cache_full_stat_dorado.pdf}
%}
\caption{The storage system with rule-based cache capacity overload control method is with shorter duration of cache being full, compared to one without overload control method. To align with Figure~\ref{fig:withandwithout}, the workload with duration of 3000 seconds is split up into 10 equal-length time slots. In each subfigure, the horizontal axis presents 10 sampled time slots and the vertical axis represents the duration of cache being full of storage system with/without cache capacity overload control method in Linux and OceanStor.} 

\label{fig:cache full duration}
\vspace{-1em}
\end{figure}
\fi

\subsection{Storage System}
\label{subsec:BGD}

\begin{figure}%\vspace{-1.6em}
\centering
%\subfigure[Comparison between systems with/without cache overload method in Linux. ]{
%\includegraphics[width=3.2in]{./fig/preliminary_linux.pdf}
%}
%\quad
%\subfigure[Comparison between systems with/without cache overload method in OceanStor.]
{
\includegraphics[width=3.0in]{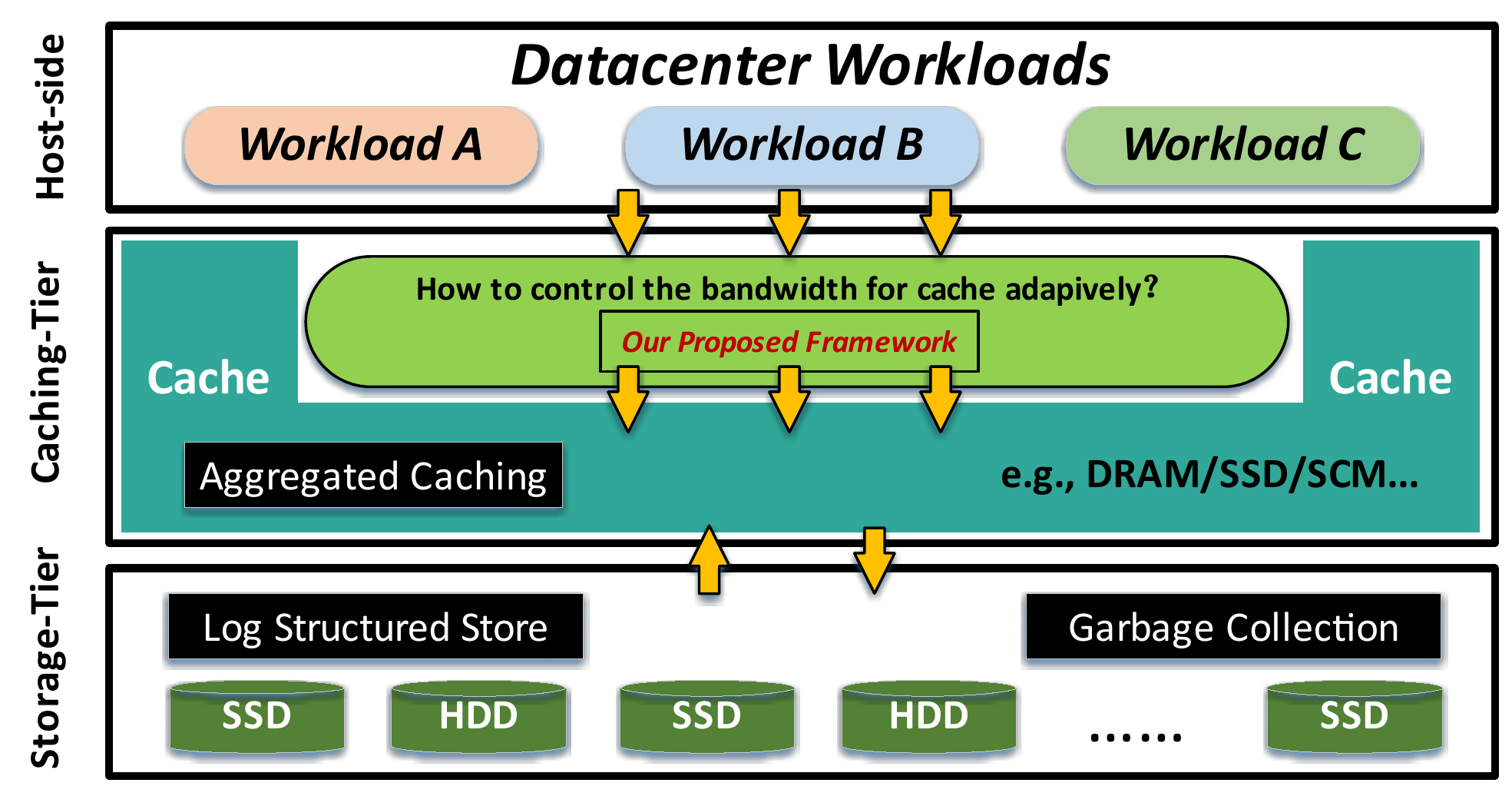}
}
\caption{The commonly-used storage architecture.}
\label{fig:OS}
\end{figure}

\iffalse
\begin{figure}\vspace{-1em}
\centering
\subfigure{
\includegraphics[width=3.0in]{./fig/linux_noCtrl_vs_base4.pdf}
}
\quad
\subfigure{
\includegraphics[width=3.0in]{./fig/dorado_noCtrl_vs_base4.pdf}
}
\vspace{-0.6em}
\caption{The comparison results for the system without overload control method, the rule-based and the learned framework.}
\label{fig:withandwithout}
\vspace{-1.3em}
\end{figure}
\fi

%\subsubsection{System Overview}
%\label{subsec:BGD}
In Figure~\ref{fig:OS}, a storage system typically consists of host-side servers, caching-tier and storage tier~\cite{Ahmadian2019, KanWu2021}. In general, host-side is comprised of general-purpose servers, which issue I/O requests to the caching-tier. Next, these I/O requests are first served by the caching-tier, and then write requests are persisted in the storage tier. 

In the caching tier, there are mainly two types of I/O requests. One is from host-side servers and the other is generated from prefetching hot data of the storage tier. Each I/O request is associated with data block numbers to access. Within a time period, a bandwidth control to host-side servers is applied, by which the total size of data blocks accessed by I/O requests cannot exceeds a given threshold. Thus, with the band control, not every I/O request from host-side servers is allowed into the caching tier, and instead it may be blocked in a network queue to be postponed to next time period, which might affect the \textit{tail latency} of the whole system. Meanwhile, the data stored in caching tier might be destaged to the storage tier or needs to be evicted out. Similarly, there is a bandwidth control for data destaging and eviction, called flush bandwidth, that control the maximum throughput for the data stored in the caching tier to be destaged or evicted to the storage tier. Note that it is hard to control the flush bandwidth, because the value of flush bandwidth might be affected by a group of background tasks (such as aggregated caching, garbage collection, etc.) and their interactions (such as resource contention)~\cite{hao2020linnos}.

The caching tier has a fixed storage capacity, which is determined by hardware. When the free space of the caching tier is lower than a given threshold, the system throughput will drop down abruptly as less bandwidth can be provided for caching data. Thus, we need to adaptively control the bandwidth to avoid the cache to be full. In this way, the throughput can be maintained stably (i.e., low \textit{throughput jitter}). On the other hand, we should maximally utilize all available bandwidth, so the cache can serve as many I/O requests (\textit{throughput}) as possible. It is a challenge for how to maximally utilize all bandwidth while prevent cache overloading.

% Among them, cache-tier plays the most significant role since it determines the overall performance of the system to some extent~\cite{paschos2019cache}.
% In performance enhancement of storage system, cache overload problem might draw more attention than other ones. Because it is highly likely to cause performance degradation if one cannot appropriately deal with the cache overload problem. Therefore, our goal is how to control the bandwidth for cache adaptively to prevent cache overload. In the following, the problem will be formally defined. 

% which might cause performance deterioration, e.g., high throughput jitter and tail latency. 

\subsection{Problem Formulation}
\label{subsec:PDC}

The bandwidth control process and cache overloading are described above. Here we formalize them in a mathematical way. For each time period $t$, we denote the throughput bandwidth by $I_t$. The flush bandwidth is represented by $O_t$. The maximum capacity of the caching tier is denoted by $C$. The watermark level $W_t$ of the caching tier is defined as follows:
% \begin{footnotesize}
\begin{equation}
    \label{eq: W_t}
    W_t = \max (0, \sum_{t'=0}^{t} (I_{t'} - O_{t'})/C)
\end{equation}
% \end{footnotesize}
where $W_{t}$ represents how much capacity has been occupied until the time period $t$. Now we can formally define the cache overloading as follows:
\begin{definition}[Cache Overloading]
The throughput bandwidth $I_t$ drops down abruptly when the watermark level $W_t$ exceeds a given threshold $\overline{W}$, i.e., 
\begin{equation}
     I_{t} \to 0 |_{W_t \ge \overline{W}}
    \label{eq: cop}
\end{equation}
\end{definition}

Our goal is to prevent cache overloading by controlling the throughput bandwidth ($I_t$), while at the same time utilizing all available bandwidth to maximize the throughput and minimizing the throughput jitter. We formulated this as follows: we would like to maintain the throughput as maximized as possible and achieve lower throughput jitter, which are formalized as:
\iffalse
\noindent\textbf{Objective}
\begin{footnotesize}
\begin{equation}
     \max \quad \sum_{t=0}^T I_t
     \label{eq: obj_max}
\end{equation}
\begin{equation}
     \min \quad \sqrt{\frac{\sum_{t = 0}^{T}(I_t - \overline{I_t})^2}{T}}/ \overline{I_t}, 
    %  \sum_{t=1}^T (I_t - I_{t-1})^2
     \label{eq: obj_stable}
\end{equation}
\end{footnotesize}
\fi

\begin{equation}
     \max \sum_{t=0}^T I_t \quad\quad
     \label{eq: obj_max}
\end{equation}

\begin{equation}
     \min \sqrt{\frac{\sum_{t = 0}^{T}(I_t - \overline{I_t})^2}{T}}/ \overline{I_t} \quad\quad\quad 
    %  \sum_{t=1}^T (I_t - I_{t-1})^2
     \label{eq: obj_stable}
\end{equation}

\noindent\textit{s.t.}

%\noindent\textbf{Constraint}
\iffalse
\begin{footnotesize}
\begin{equation}
    \label{eq: W_t_1}
    W_t = \max (0, \sum_{t'=0}^{t} (I_{t'} - O_{t'})/C),\quad
\end{equation}
\begin{equation}
     \underline{W} \leq W_t \leq \overline{W}\quad
     \label{eq: constr_waterlevel_threshold}
\end{equation}
\begin{equation}
     O_t \sim \mathcal{P}(t|\xi)\quad
     \label{eq: constr_o_t}
\end{equation}
\end{footnotesize}
\fi

% \begin{footnotesize}
% \begin{tabular}{rr}
    %  \begin{equation}
    % \label{eq: W_t_1}
    % W_t = \max (0, \sum_{t'=0}^{t} (I_{t'} - O_{t'})/C),
% \end{equation}
\begin{equation}
    W_t \leq \overline{W} \quad\quad\quad
     \label{eq: constr_waterlevel_threshold}
\end{equation}
\begin{equation}
     \quad\quad\quad\quad O_t \sim \mathcal{P}(t|\xi) \quad\quad\quad\quad\quad
     \label{eq: constr_o_t}
\end{equation}
% \end{tabular}
% \end{footnotesize}

\noindent where $T$ is the duration of the bandwidth control process and $\overline{I_t}$ represents the average throughput of the process. Eq.(\ref{eq: obj_max}) represents the cumulative throughput should be maximized. Eq.(\ref{eq: obj_stable}) denotes the throughput jitter should be minimized if the cache overloading could be avoided.
Empirically, the tail latency of the system will be optimized once the two above objectives are achieved. Eq.(\ref{eq: constr_waterlevel_threshold}) constrains the watermark level to be less than threshold so as to avoid the cache overloading.
As described above, the flush bandwidth can be affected by multiple background tasks. Thus we treat the flush bandwidth as a random variable subject to time period $t$ and other system factors $\xi$, as shown in Eq.(\ref{eq: constr_o_t}).
\vspace{-1.0em}

\subsubsection{Two Rule-based Methods}
\label{subsec:BGD}
%To solve the cache overload problem above, some ruled based methods were proposed. Two representative studies are: `bypass' in Linux platform and the method we referred  as `\CoTo{}' in Huawei storage product OceanStor.
\header{`bypass'} It refers to that if the I/O requests into the cache become congested, which are detected by latency of the I/O requests exceeding a predefined threshold, the system will directly forward these I/O requests to the storage tier. Otherwise, these I/O requests will enter the cache. 
\header{`\CoTo{}'}
It is a finite state machine, where the cache state is depicted by the discretized watermark level such as \{High, Mid, Low\}. 
For any given time point $t$, \CoTo takes the state as input and outputs the predefined action (such as \{Decrease, Hold, Increase\}) for the current bandwidth of I/O requests into the cache. 
For example, \CoTo opts for increasing a given amount of the bandwidth of I/O requests into the cache when the watermark level is low, or vice versa. %Note that this bandwidth limiting method implemented by token bucket is described in Section~\ref{subsec:overview}. For a given time point $t$, the watermark level $W_t$ is mapped into one of the discrete states $S_t \in \mathbf{S}$ ($\mathbf{S}$=\{High, Mid, Low\}) according to predefined range intervals. The next time point of bandwidth $I_{t+1}$ is calculated by \CoTo as
%\begin{equation}
%    $I_{t+1} = (1+\alpha_t)\times I_t$,
%\end{equation}
%\begin{equation}
%\label{eq1}
%$\alpha_t$ =
%\{\begin{array}{rcl}
%$(\frac{W_t-W_{t-1}}{W_{t-1}}\times \frac{|I_t-O_t|}{O_t})/-r/r/0$  while ${S_t \neq S_{t-1}/= High/Low/others,}$ 
%\end{array} \right.
%\end{equation}
%where $r$ is the base value for adjustment rate. 
%$\frac{W_t-W_{t-1}}{W_{t-1}}$ indicates the adjustment direction (a negative value represents an increase and a positive value represents a decrease).
%$\frac{|I_t-O_t|}{O_t}$ is the change rate.
%Extensive experiments have been performed, which suggest that the best configurations are $r=5\%$, $\mathbf{S} = \{Low \in (0, 20), Mid\in [20, 80), High\in (80, 100]\}$. 

We summarized the limitations of these two rule-based methods as follows: \textbf{(1)} they are \textbf{\textit{highly customized and require elaborately handcrafted design}} which cannot be generalized to time-varying external demands and internal issues. \textbf{(2)} they are not able to continuously ameliorate the cache being full over time because there is \textbf{\textit{no feedback mechanism}} for these rule-based methods to perceive or learn from the change of storage system state and workload. \textbf{(3)} The rule-based method \CoTo \textbf{\textit{using the watermark level alone}} (`bypass' \textbf{\textit{only measures by the latency of I/O request}}) is not sufficient for a comprehensive measure of the system state for bandwidth control and will lead to some unnecessary or improper bandwidth adjustment operations. 

\begin{figure}%\vspace{-2em}
\centering
{
\includegraphics[width=3.5in]{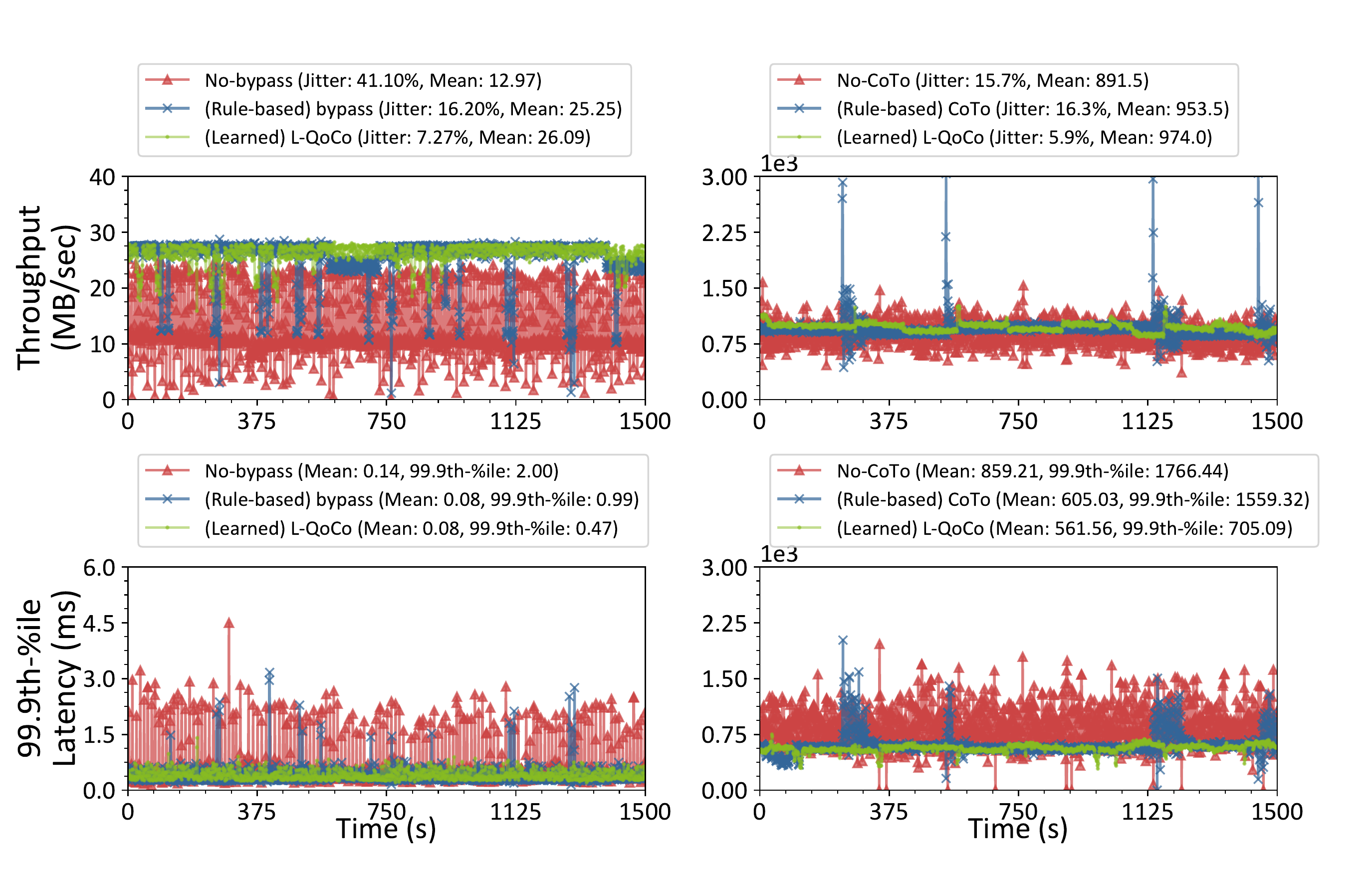}
}
\vspace{-1em}
\caption{The comparison results for the system without overloading control method, the rule-based and the learned framework. 
Comparison of the throughput and P99.9 latency between `bypass' and \LQoCo on a Linux storage server and between \CoTo and \LQoCo on a Huawei OceanStor storage system.}
\label{fig:withandwithout}
%\vspace{-.3em}
\end{figure}

\vspace{-0.6em}
\subsection{Motivation}
\label{subsec:BGD}

We have conducted several preliminary experiments on two real storage systems, Linux and OceanStor (See Section~\ref{meth} for detailed experimental settings). Specifically, we first compare \LQoCo with Linux Bcache `bypass' (as a representative for cache bypassing) on a Linux storage server, and then \LQoCo with \CoTo (as a representative for finite-state-machine based heuristics). The workload is S-3 (See Section~\ref{meth} for detailed description of the benchmark workload). In Figure~\ref{fig:withandwithout}, we show the experimental results for a 1500s window (similar trends can be observed in other parts).

As shown in Figure~\ref{fig:withandwithout}, it can be observed that both the two rule-based methods `bypass' and \CoTo perform better compared with the original systems in terms of the throughput and P99.9 latency. However, with both `bypass' and \CoTo, there are clearly throughput and P99.9 latency spikes, demonstrating their inability to quickly adapt to the complex storage systems. In contrast, our proposed learned framework \LQoCo outperforms `bypass' and \CoTo with much more stable throughput and better P99.9 latency.

\begin{figure}[t]%\vspace{-2em}
\centering
\includegraphics[width=3.in]{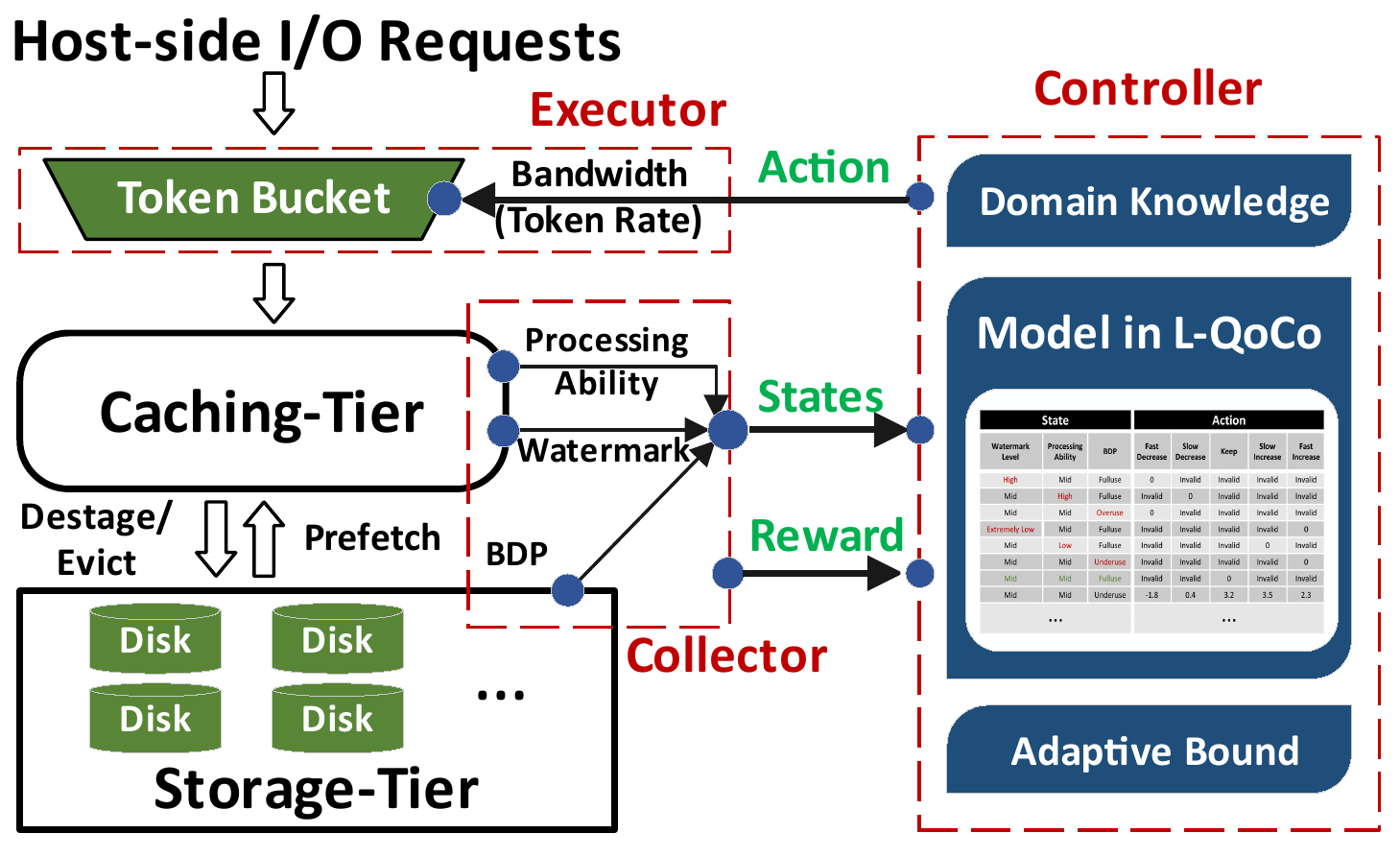}
\caption{The overview of \LQoCo. }
\label{fig:overview}
\end{figure}

\vspace{-1.16em}
\section{The Proposed Technique} 
\vspace{-0.3em}

\subsection{Overview}
\label{subsec:overview}
\vspace{-0.3em}

%Keeping the above principles in mind, \LQoCo is proposed. 
As shown in Figure~\ref{fig:overview},
\LQoCo consists of three main components, i.e., state collector, controller and executor. In \LQoCo, we adopt Q-learning as the learning algorithm. Q-Learning~\cite{watkins1992q} is one of the most classic RL algorithms. It has been verified that Q-learning performs well in discrete state space~\cite{osmankovic2011implementation}, which is in accordance with our lightweight requirement.   

%The state collector periodically collects kinds of system information. This information will be taken as the input into the controller. A recommended bandwidth limiting value is determined by the controller. Then according to the recommended bandwidth limiting value, an executor will adjust the bandwidth (i.e., the token rate) of the cache.

% mainly consists of a state collector and an RL-based controller.
% The former collects system states periodically and the latter takes as input the system states and decides a bandwidth limiting value using the RL technology.
% We first discuss how to apply RL to simulate the trial-and-error method for performing an adaptive bandwidth limiting to overcome the drawbacks of our previously proposed \CoTo (RL-based Controller in Section~\ref{RL4L}).
% Then we introduce Q-learning in our \LQoCo (Section~\ref{QL}).
% Besides, we describe the effective reward function (Section~\ref{Reward}) design in \LQoCo and the domain knowledge for model acceleration (Section~\ref{Acceleration}).
% Lastly, we present a safeguard bound strategy (Section~\ref{Safeguard}) for safe exploration in \LQoCo.

\header{Collector} To make the controller concentrate on policy learning and inference, lots of `dirty' works are moved into the collector. Specifically, the collector is to collect raw data and to preprocess these data into various statistics (e.g. $W_t$, $I_t$, $O_t$). Then these statistics are taken as the inputs of the controller.

\header{Controller} The controller takes the system state from the collector as the inputs and calculates an appropriate bandwidth $I_t$. Within the controller, Q-learning is adopted for decision making in the fashion of online learning, which makes \LQoCo adaptive to time-varying workloads. Besides, for high availability and robustness, an adaptive bound strategy is judiciously devised.

% such as watermark level, the number of I/O requests out of/into the cache, etc (detailed data collected by collector see Section~\ref{subsec:overview}). Then it also needs to preprocess these raw data into predefined statistics according to eq.(\ref{eq: W_t}). These statistics are then taken as input into the controller.

% The state collector is designed primarily for system states collecting and processing. It collects curtain raw data from the storage system, and then it processes the data to fit the model input data format.
\header{Executor} For bandwidth control, we utilize a token-bucket mechanism~\cite{kramer2019cooperative}. Specifically, it adds a given number of tokens into the bucket according to the recommended bandwidth $I_t$ obtained from the controller. There is a predifined conversion formula between the number of tokens and bandwidth. %For example, within Huawei OceanStor, one byte of write I/O request consumes one unit of token. %Readers of interest can refer to~\cite{kramer2019cooperative, faheemperformance}.
% In other words, the number of added token is equal to the bandwidth.}

% into which tokens are added at the same rate as the recommended bandwidth from the RL-based controller. 
% When a writing I/O request arrives at the executor, if there is sufficient tokens, it will consume tokens by its size and the data will be written into the cache. If the size of remaining tokens is less than that of the writing data, this data can only wait inside the executor until more tokens are added.
% As a result, the executor is able to determine the maximum bandwidth of I/O request into the cache.
% 推荐值给qos之后有个分配令牌的机制，这块需要写令牌桶的机制吗？我个人想法是不写详细了，因为不是重点工作，但这样的话感觉可以不专门起一个标题介绍这块了？
\vspace{-1.3em}
\subsection{Reinforcement Learning in \LQoCo}

\subsubsection{\textit{Markov} decision process}
%mapping,Reward Function
%Q-Learning in \LQoCo
\label{RL4L}

% The main challenge of using RL in \LQoCo is to map an overload control mechanism for cache in storage systems to appropriate actions in RL.%TODO: moved to Challenges
% Figure~\ref{fig:QoCo}~(a) shows an interaction diagram of the six key elements in RL and their correspondence between the six elements and the overload control mechanism for cache in the storage system.

% \textcolor{black}{We formulate the cache capacity overload problem as an \textit{Markov} decision process, which could be understood as a mapping relation between elements in RL and components of the cache capacity overload problem. 
% To intuitively show the mapping relation, Figure~\ref{fig:QoCo}~(a) is given.
% }

% \begin{figure}[t]\vspace{-0em}
% \centering
% \includegraphics[width=3.4in]{./fig/RL-QoCo.pdf}
% \caption{(a) The relation between RL elements and cache overload control framework. (b) Abstracting the system states to several blocks in game Maze for better understand.}
% \label{fig:QoCo}
% \end{figure}

% To intuitively show the proposed \LQoCo, Figure~\ref{fig:QoCo}~(a) is given to present the mapping relation between elements of reinforcement learning and components of overload control for cache in the storage system.}

%\begin{figure}[t]
%\centering
%\includegraphics[width=3.36in]{./fig/QoCo.PNG}
%\caption{The correspondence between RL elements and overload control mechanism for cache in storage system.
%}
%\label{fig:QoCo}
%\end{figure}

% \header{Environment \& Agent} 

Reinforcement learning is to achieve a goal by learning from interactions. The learner or decision-maker is named the agent. The thing it interacts with, comprising everything outside the agent, is called the environment. In this paper, the agent refers to the controller of \LQoCo and the environment represents the cache in the storage system. Based on a policy, the agent acts on the environment (i.e., determines the throughput bandwidth) according to real-time state information collected from the environment. The policy is updated (learned) by reward calculated from the state transition. In the following, we first formulate the control process as an \textit{Markov} decision process (MDP). Then the proposed algorithm is introduced. 
% , which is supposed to reveal the state transition of the system (via the collector) according to the action performed on it (via the executor). 

% The agent aims to learn an optimal bandwidth control policy via interacting with the environment.

\header{Objective of RL} The goal of \LQoCo is to learn an optimal bandwidth control policy that can maximize the throughput bandwidth while preventing cache overloading. 

% 1) maintain throughput as high as possible and meanwhile 2) can keep the throughput jitter and tail latency as low as possible. 

\header{State} For a given time period $t$, the state $S_t = (W_t, P_t, B_t)\in\mathbb{R}^3$ is utilized to describe the internal status of the cache. Specifically, the watermark level $W_t$ is defined in Section~\ref{subsec:PDC}; $P_t$ stands for the I/O request processing ability of the cache, which is calculated by $P_t = \frac{I_t-O_t}{O_t}$; and $B_t$ is the bandwidth-delay product (BDP)~\cite{Cardwell2017}, which estimates the maximum value of the flush bandwidth.
% Empirically, we found that one of the most important factors determining the throughput of the whole system is the maximum capability of receiving I/O requests in the storage backend. 
% However, this capability is affected by varieties of factors (such as garbage collection, disk failure, data deduplication and unbalanced resource scheduling~\cite{kougkas2019labios}), whose fluctuation will result in congestion of I/O request.
% Besides, we use  to estimate  $B_t$ is calculated as $B_t = O_t\times d_{avg}$, where $d_{avg}$ is the average latency of all I/O requests that the cache destages to the storage backend in a predefined period (we set the period as $5$ seconds).
% The larger $B_t$ is, the lower the maximum capability of receiving I/O requests in the storage backend is and vice versa. 
% In a word, $S_t$ is formulated as $S_t = (W_t, P_t, B_t)\in\mathbb{R}^3$. 
To make the RL solution lightweight, $S_t$ is discretized. We denote the discretized state by $\hat{S}_t = (\hat{W}_t, \hat{P}_t, \hat{B}_t)$. 
Correspondingly, $W_t$ is discretized as $\hat{W_t} \in \mathbf{W}=$  \{ExtremelyLow, Low, Mid, High\}; $P_t$ is discretized as $\hat{P}_t \in \mathbf{P}=$ \{Low, Mid, High\}; and $B_t$ is discretized as $\hat{B}_t \in \mathbf{B}= $ \{Overuse, Fulluse, Underuse\}. 

\header{Action} For a given discretized state $\hat{S}_t$, the action $a_t\in [0,1]$ refers to the adjustment rate of the throughput bandwidth. Specifically, $I_{t+1}$ is determined by $I_{t+1}=(1+a_t)\times I_t$. The adjustment rate $a_t$ is also discretized as \{SlowDecrease(-1\%), FastDecrease(-3\%), Keep(0\%), SlowIncrease(1\%), FastIncrease(3\%)\}, denoted by $\hat{a}_t$. Note that except for `Keep' action, the values of the rest are hyperparameter.

\header{State transition} Once \LQoCo takes an action (i.e., determines $I_{t+1}$) for current cache state $\hat{S}_t$, there is a state transition from $\hat{S}_t$ to $\hat{S}_{t+1}$. The transition is stochastic, denoted by $p(\hat{S}_{t+1}|\hat{S}_t, \hat{a}_t)$.
The stochastic comes from many perspectives such as time-varying workloads, complex background tasks (such as garbage collection, aggregated caching), etc.

\begin{table}[t]

\begin{center}%\vspace{-2em}
\caption{Categorization of state based on domain knowledge.}
\scalebox{0.80}
{
\begin{tabular}{|c|c|c|}

\hline
Name & Notation & Definition \\
\hline
Better & $\mathbf{S}_b$ & $\{\hat{S}_t|$ $\hat{W}_t=$ Mid $\land$ $\hat{P_t}=$ Mid $\land$ $\hat{B_t}=$ Fulluse $\}$\\
\hline
Worse & $\mathbf{S}_w$ & $\{\hat{S}_t|$ $\hat{W}_t=$ High $\land$ $\hat{P_t}\in$ \{Low, High\} $\land$ $\hat{B_t}=$ Overuse $\}$\\
\hline
General & $\mathbf{S}_g$ & $\{\hat{S}_t|\hat{S}_t\notin \mathbf{S}_b$ and $ \hat{S}_t\notin \mathbf{S}_w \}$ \\
\hline
\end{tabular}
}
\label{tab: state categorization}
\end{center}
% \vspace{-1em}
\end{table}

\begin{table}[h]\vspace{-1em}
\begin{center}
\caption{Reward assigned to each dimension of discrete state.}
\scalebox{0.80}
{
\begin{tabular}{|c|c|c|c|c|c|}

\hline
$\hat{W}_t$ & Value & $\hat{P}_t$ & Value & $\hat{B}_t$ & Value\\
\hline
ExtremelyLow & 0 & \textcolor{red}{Low} &\textcolor{red}{-1} & Underuse & 0\\
\hline
Low & 0 & Mid & 0 & \textcolor{blue}{Fulluse} & \textcolor{blue}{1}\\
\hline
    \textcolor{blue}{Mid} & \textcolor{blue}{1} & \textcolor{red}{High} & \textcolor{red}{-1} & \textcolor{red}{Overuse} & \textcolor{red}{-1}\\
\hline
\textcolor{red}{High} & \textcolor{red}{-1} & -- & -- & -- & -- \\
\hline
\end{tabular}
}
\label{state value}
\end{center}
\vspace{-1em}
\end{table}

\header{Reward} The reward $R_t$ is calculated from the state transition given action $a_t$, i.e., $R_t = R_{\hat{a}_t}(\hat{S}_{t+1}, \hat{S}_t)$ , which is the signal for the agent to learn its policy. To devise an effective reward function, we first categorize the above discretized states as three classes, namely `Better', `Worse' and `General'. The state category is listed in Table~\ref{tab: state categorization}. The different rewards are assigned to these categorized states according to domain knowledge. 
A storage system with high availability should present better system states and rapidly escape from the worse states. Thus, those states $\hat{S}_t \in \mathbf{S}_b$ deserve positive reward and those states $\hat{S}_t \in \mathbf{S}_w$ are supposed to be given negative reward. The rest of states $\hat{S}_t \in \mathbf{S}_g$ just get the intermediate level of reward. Based on above, we design a value assignment for each dimension of discrete state, which is summarized in Table~\ref{state value}. Then the reward function is given as follows:
% \begin{footnotesize}
\begin{equation}
    \begin{split}
    R_t = (\frac{\hat{W}_{t+1}^2}{(|\mathbf{W}| - 1)^2} - 1)\times f_W \\
     +(\frac{\hat{P}_{t+1}^2}{(|\mathbf{P}| - 1)^2} - 1)\times f_P \\
    + (\frac{\hat{B}_{t+1}^2}{(|\mathbf{B}| - 1)^2} - 1)\times f_B
    \end{split}
    \label{eq: reward1}
\end{equation}
% \end{footnotesize}
% \begin{footnotesize}
\begin{equation}
    f_W+f_P+f_B = 1
    \label{eq: reward2}
\end{equation}
% \end{footnotesize}
where $f_W, f_P$ and $f_B$ represent the weight of each dimension of a discrete state, respectively, and their sum should be equal to one. The weight value is also a hyperparameter. The reward function is independent of the hardware environment and workload changes, but only depends on the states.
Thus, it is relatively easy to adapt the reward function to other definition of states. 

\vspace{-0.8em}
\subsection{Enhancements for Learning in \LQoCo}

\subsubsection{Acceleration of policy learning using domain knowledge}
\label{Acceleration}

Most of RL algorithms are sample-inefficient~\cite{pietquin2011sample}. Besides, the convergence speed of the RL model depends on both the size of the states and action space. %For example, AlphaGo~\cite{Wang2016}, the first computer program to defeat a world champion at the game of Go, was trained in millions of CPU/GPU hours. 
Even for relatively simple applications, the training time varies from minutes to hours or even days, which is undesired in storage systems.
% In storage system, such huge amount of required computing resource is impractical for providing for training RL model in \LQoCo.
Thus we resort to using domain knowledge to narrow the learning space.

%\vspace{-1.2em}
% \algnewcommand{\IIf}[1]{\State\algorithmicif\ #1\ \algorithmicthen}
% \algnewcommand{\EndIIf}{\unskip\ \algorithmicend\ \algorithmicif}

% \begin{footnotesize}\vspace{-0.68em}
\begin{algorithm}[h]

%\algsetup{linenosize=\tiny}\scriptsize
	\caption{Adaptive bound mechanism in \LQoCo{}}
	\begin{algorithmic}[1] %每行显示行号
		\STATE \textbf{Input:} 1) initial lower bound $lb$ and initial upper bound $ub$; 2) threshold $N$ and $v$; 3) update factor $\sigma$ and $\delta$
		\STATE \textbf{Initialize:} 1) count of system keeping at better state $C_b=0$; 2) extent of $I_t$ violating upper and lower bound $V_{ub}, V_{lb} =0$; 3) historical queue $Q_I=\{\}$; 4) timestamp $t=0$
		\WHILE{True}
		\STATE Get recommended bandwidth $I_t$ from learned policy and store $I_t$ into $Q_I$
		\IF{$S_t$ is in better state}
		\STATE $C_b=C_b+1$ \tcp{updates the count of better state}
        \ENDIF
        \IF{$I_t > ub$}  
        \STATE $V_{ub} = (V_{ub} - 1) \times \delta + 1$ 
        \ENDIF
        \IF{$I_t < lb$} 
        \STATE $V_{lb} = (V_{lb} - 1) \times \delta + 1$ 
        \ENDIF %\tcp{update violation extent} 
% 		\tcp{Bound shrinkage}
		\IF{$C_b \ge N$}
		\STATE $\overline{I}=\frac{1}{N}\sum_{k=0}^{N-1} I_{t-k}$ %\tcp{the average of previous $I_t$}
		\STATE $lb=\overline{I}\times (1-\sigma)$ and $ub=\overline{I}\times (1+\sigma)$ %\tcp{update the lower and upper bound}
		\STATE $C_b=0$ \tcp{reset the count}
        \ENDIF
        % \tcp{One-side bound update}
        \IF{$V_{lb}>v$} 
        \STATE $lb=I_{t-1}$ 
        \STATE $V_{lb}=0$ 
        \ENDIF %\tcp{update bound and reset violation extent}
        \IF{$V_{ub}>v$} 
        \STATE $ub=I_{t-1}$ 
        \STATE $V_{ub}=0$ 
        \ENDIF%\tcp{update bound and reset violation extent}
		\STATE $t=t+1$ \tcp{timestamp +1}
		\ENDWHILE
	\end{algorithmic}
	\label{alg: adaptive_bound}
\end{algorithm}
% \end{footnotesize}

% Integrating domain knowledge inting dto RL might significantly reduce the time of policy training and meanwhile increase the reliability, interpretability and robustness of the RL policy~\cite{Abhinav2018}. To this end, we introduce useful domain knowledge of cache overload control into \LQoCo.
% \vspace{-1.em}
\header{Safe action to extreme states} Some discrete states are recognized as `extreme' states. When the cache reaches these extreme states, \LQoCo is supposed to instantly take a safe action instead of following its learned policy, since it is highly likely that the storage system becomes extremely worse if \LQoCo does not immediately takes the corresponding safe action. Two extreme states and corresponding safe actions are defined in Table~\ref{tab: extreme state}. It should be noted that the policy is not updated if the system enters into extreme states.

\begin{table}[t]
\begin{center}
\caption{Definition of extreme states and safe actions.}
\scalebox{0.70}
{
\begin{tabular}{|c|c|}

\hline
Extreme state & Safe action \\
\hline
  $\{\hat{S}_t |$ $\hat{W}_t$=High $\lor \hat{P}_t$ = High $\lor \hat{B}_t$=Overuse $\}$ & SlowDecrease or FastDecrease \\
  \hline
  $\{\hat{S}_t |$ $\hat{W}_t$=ExtremelyLow $\lor \hat{P}_t$ = Low $\lor \hat{B}_t$=Underuse $\}$ & SlowIncrease or FastIncrease \\
  
\hline
\end{tabular}
}
\label{tab: extreme state}
\end{center}

\end{table}

\header{Fine tuning in better states} Although the system is constantly in `Better' state, it does not mean that we have nothing to do. Since the flush bandwidth (i.e., $O_t$) is time-varying, we still have to slightly tune the throughput bandwidth $I_t$ when the storage system keeps at the `Better' state. For example, the bandwidth is supposed to be finely adjusted (e.g., $0.1\%$ of $a_t$) based on the changing trend of the watermark level (when $W_t$ lightly decreases, the $I_t$ should slightly increase, and vice versa). In this way, the fluctuation of $O_t$ could be offset to some extent.

% We first define some boundary system states. The agent will directly take a safe action without following its policy (i.e., Q table) once the environment enters these boundary states. For example, when the storage system reached such states (the watermark level $W_t$ becomes High or cache process ability $P_t$ is High or $B_t$ is Overuse), the action of SlowDecrease or FastDecrease will be directly adopted regardless of the action outputted from the Q table. Similarly, when $W_t$ becomes ExtremelyLow or $P_t$ is low or $B_t$  is Underuse, the action of SlowIncrease or FastIncrease will be taken immediately.

% we will not employ ``decrease'' actions.

% we employ ``decrease'' actions instead of ``increase'' actions output by \LQoCo, because ``increase'' actions may lead to writing the cache full which degrade the overall performance.

% it is difficult to keep the storage system in better states through a constant bandwidth limiting. Therefore, we should finely adjust (e.g., 0.1\% step size) the token rate for the token bucket based on the trend of the watermark change (the watermark level lightly decreases, the bandwidth slightly increase, and vice versa) to adapt to the environment changing instead of drastic adjustment when the storage system enters the better states. Note that the \LQoCo will not update Q table under these situations.
% \vspace{-1em}
\subsubsection{Robustness improvement with adaptive bound mechanism}
\label{Safeguard}

% The extreme states and corresponding actions have been defined in the last subsection, which takes effect only when extreme state occurs. However, there is no any restriction on bandwidth recommended by Q table. 

Without any restriction, the learned policy in \LQoCo might lead the system into an unexpected state or even system collapse. For instance, extremely high $I_t$ might easily make the cache full and further lead to congestion of I/O requests; extremely low one cannot make full use of the bandwidth resource, thus degrading the overall performance of storage systems. 
% In additions, considering that the workloads are time-varying in the storage system, the learning algorithm might fail to converge before the workload starts to change. 
% Thus, it is required that we could prevent the storage system from entering extreme state in advance. 
Hence, in \LQoCo we design an adaptive bound mechanism for bandwidth recommended by the learned policy. Specifically, additional lower and upper bound ($lb$, $ub$) are constantly maintained in \LQoCo{}, where $I_t$ is supposed to be within the range between the lower and upper bounds. Otherwise, $I_t$ will not be executed in the system. Note that \LQoCo will not update the policy if the recommended bandwidth is not executed. The update method for $lb$, $ub$ is described in Algorithm~\ref{alg: adaptive_bound}.

\vspace{-0em}
\section{Evaluation}
\label{sec:ES}

%\textcolor{black}{Next, the methodologies are first presented. Then with the methodologies, overall evaluation and evaluation for more intuitions. }

%, the proposed learning-based framework \LQoCo is compared with the rule-based ones (\CoTo, `bypass') and no control mechanisms (\NoCoTo, `No-bypass') under different platforms and types of workloads to verify the effectiveness and efficiency of \LQoCo.}
%-------------------------------------------------------------------------------
%\vspace{-0.6em}
\subsection{Experimental Setup and Implementation}
\label{meth}

\header{For a comprehensive evaluation, we have conducted experiments with the following two platforms}
\textbf{(1) Linux:} 
The first platform is a Linux storage server equipped with a 6-core 3.2 GHz Intel(R) Xeon(R) Gold 6278C CPU, a cache tier (16GB of RAM), and a hierarchical  storage tier (2GB SSD and 10GB HDD). The server is installed with Linux 5.8.0 and Ext4 file system, and the LRU cache replacement algorithm and the writeback cache write policy are utilized in page cache management. 
\textbf{(2) Huawei OceanStor:} 
The second platform is a Huawei OceanStor storage server (Dorado 3000 V6\footnote{https://e.huawei.com/en/products/storage/all-flash-storage/dorado-3000-v6}) equipped with two 24-core 2.6GHz  Kunpeng-920 CPUs, a cache tier (6*32 GB of RAM) and a storage tier (25*0.96TB SSDs). The server is installed with Huawei OceanStor software for cache and storage management. Note that we used the different hardware mediums as a cache for two storage systems (RAM for OceanStor and SSD for Linux) 
to identify the generalization of our proposed framework. Besides, we also employed same cache write policy in these two platforms for fair comparison.

%OceanStor Dorado 3000 V6 \footnote{https://e.huawei.com/en/products/storage/all-flash-storage/dorado-3000-v6} storage servers with 24-core 2.6GHz CPU*2, 32GB RAM*6 and 0.96TB SSD*25 whose RAM is regarded as cache and SSD is regarded as storage tiering. Note that we used the different hardware mediums as a cache for two storage systems %(RAM for OceanStor and SSD for Linux) to identify the generalization of our proposed framework. 

\header{Implementation of two rule-based methods}
\textbf{(1) `bypass' in Linux:} We directly turn on the knob `bypass' in Bcache on Linux platform.  
\textbf{(2) `\CoTo{}' in OceanStor:} For a given time point $t$, the watermark level $W_t$ is mapped into one of the discrete states $S_t \in \mathbf{S}$ ($\mathbf{S}$=\{High, Mid, Low\}) according to the predefined time interval. The next time point of bandwidth $I_{t+1}$ is calculated by \CoTo as follows:
% \begin{footnotesize}
\begin{equation}
    I_{t+1} = (1+\alpha_t)\times I_t
\end{equation}

\begin{equation}
\label{eq1}
\alpha_t =\left
\{\begin{array}{rcl}
\frac{W_t-W_{t-1}}{|\mathbf{S}|}\times \frac{|I_t-O_t|}{O_t} & & {S_t \neq S_{t-1} }\\
-r & & {S_t = High}\\
r & & {S_t = Low}\\
0 & & {others}
\end{array} \right.
\end{equation}
% \end{footnotesize}
where $r$ is the base value for
the adjustment rate. $|\mathbf{S}|$ is the size of the enumerated value $\mathbf{S}$ (e.g. $|\mathbf{S}|$ = 3 when $\mathbf{S}$ = \{0, 1, 2\}). 
$\frac{W_t-W_{t-1}}{|\mathbf{S}|}$ indicates the adjustment direction (a negative value represents an decrease and a positive value represents a increase).
$\frac{|I_t-O_t|}{O_t}$ is the change rate.
Extensive experiments have been performed, which suggest that the best configurations are $r=5\%$, $\mathbf{S}=\{Low \in (0, 20), Mid\in [20, 80), High\in (80, 100]\}$.

\header{Implementation of \LQoCo}
\textbf{(1) \LQoCo for Linux:} if the bandwidth of I/O requests from the host-side is larger than the bandwidth value recommended by \LQoCo, the exceeding parts of I/O requests (according to the order of I/O requests) will directly enter the storage tier instead of using the predefined latency threshold in `bypass'.
\textbf{(2) \LQoCo for Huawei OceanStor:} we directly replace \CoTo (the default cache management policy in Huawei OceanStor) with \LQoCo.

\begin{table}[t!]
\caption{Description of the workload traces. }
\label{workloaddet}
\centering
\scalebox{0.90}{
\begin{tabular}{|c|c|c|c|}
\hline
\# & Block Size & I/O Size & R/W \\ \hline
S-[1,2,3,4] & 512B & 4KB & [0,30,50,70]\% \\ \cline{1-4}
S-[5,6,7] & 8KB & 32KB & [0,30,70]\% \\ \cline{1-4}
S-[8,9,10] & 8KB & 256KB & [0,30,70]\% \\ \cline{1-4}
S-[11,12,13] & 32KB & 32KB & [0,30,70]\%  \\ \cline{1-4}
S-[14,15,16] & 32KB & 256KB & [0,30,70]\%  \\\cline{1-4}
S-[17,18,19] & 8KB & 8KB & [0,30,70]\%  \\ \cline{1-4}
R-[A,B,C,D,E] & \multicolumn{3}{c|}{Real-world workloads from our customers.} \\ \cline{1-4}
		
%\textbf{Real-world} & \multicolumn{6}{c|}{\makecell[c]{R-1, R-2, R-3, R-4 and R-5 are the real-world \\workloads from the service of customers which have \\different I/O type (Random or Sequential), hybrid \\I/O size (4KB to 512KB) and different read/write ratio. }} \\

\end{tabular}
}
\end{table}

\header{Workloads}
%With few previous works about the cache capacity overload problem in storage system optimization, there is no public benchmark for performing our experiments. Therefore, 
The workloads fall into two classes, i.e., standard workload traces and real-world traces from the service of customers. %We listed the detailed properties of all the workloads we used in Table~\ref{workloaddet}. In general, they are different in the terms of I/O size, I/O type (random or sequential), the number of threads, block size and read/write ratio (write-only, read-write combination, etc.). 
For the standard benchmark workload traces, the open-source benchmarks including Filebench,
%~\footnote{https://github.com/filebench/filebench} 
FIO,
%~\footnote{https://fio.readthedocs.io/en/latest/fio\_doc.html} 
and Vdbench
%\footnote{https://www.oracle.com/downloads/server-storage/vdbench-downloads.html}
are utilized, each of which is associated with one typical business model of the users, such as database, heavy computing in AI tasks or high performance computing\cite{Tang2021}. These tools are commonly used to test and validate storage systems. %They are capable of generating I/O workloads based on parameters description, among other things, read/write mix, queue depth, block size, I/O size, thread and sequentiality with the goal of better approximating real-world workloads~\cite{JakeWires2014}. %In detail, we can tune the aforementioned parameters in those tools to obtain any type of workload. 
Table~\ref{workloaddet} lists the detailed properties of all the 19 classes of standard storage workload traces (S-1 to S-19). Besides, we collect five classes of real-world workloads (R-A to R-E), which are different in I/O type (Random or Sequential), hybrid I/O size (4KB to 512KB) and read/write ratio. %All of the real-world workloads reflect on the system performance degradation (high tail latency and throughput jitter) caused by cache capacity overload problems, from the customers. %It should be noted that we set enough (or maximum) threads for all of the standard workload traces to ensure that the cache capacity in the storage system is overloaded. 

\begin{table*}[h]%\vspace{-2em}
\centering
\caption{Performance comparison for learned framework (\LQoCo), rule-based methods (\CoTo [C] and `bypass' [B]) and the systems without control methods (\NoCoTo [NC] and `No-bypass' [NB]) on different platforms and storage workloads. %The arrow in the C/B column compares \CoTo /`bypass' with \NoCoTo/'No-bypass', arrow in the L column compares \LQoCo with \CoTo/'bypass' and \NoCoTo/`No-bypass'. 
%\LQoCo achieves better performance than \CoTo/`bypass' and \NoCoTo/`No-bypass' in almost all cases. 
}
\label{overallresult}
\scalebox{0.78}
{
\begin{tabular}{|c|c|c|c|c|c|c|c|c|c|c|c|c|c|}
\hline
\multirow{2}*{Platform} & \multirow{2}*{Workload Case} & \multicolumn{3}{c|}{Throughput (MB/s)} & \multicolumn{3}{c|}{Mean Latency (ms)} & \multicolumn{3}{c|}{99.9th-\%ile (ms)} & \multicolumn{3}{c|}{Throughput Jitter} \\ \cline{3-14}
                    ~& ~& NC/NB & C/B & \LQoCo & NC/NB & C/B & \LQoCo & NC/NB & C/B & \LQoCo & NC/NB & C/B & \LQoCo \\
%linux
\hline
\multirow{4}*{Linux} & S-1 & 22 & 24 ($\uparrow$) & 28 ($\uparrow \uparrow$) & 0.14 & 0.12 ($\downarrow$) & 0.09 ($\downarrow \downarrow$)& 1.94 & 1.65 ($\downarrow$) & 0.72 ($\downarrow \downarrow$)& 38.3\% & 39.2\% ($\textcolor{red}{\uparrow}$) & 7.2\% ($\downarrow \downarrow$)\\\cline{2-14}

~ & S-2 & 16 & 18 ($\uparrow$) & 23 ($\uparrow \uparrow$) & 0.16 & 0.08 ($\downarrow$) & 0.08 ($\downarrow \downarrow$)& 2.09 & 1.17 ($\downarrow$) & 0.57 ($\downarrow \downarrow$)& 37.9\% & 14.6\% ($\downarrow$) & 3.5\% ($\downarrow \downarrow$)\\\cline{2-14}

~ & S-3 & 13 & 16 ($\uparrow$) & 18 ($\uparrow \uparrow$) & 0.14 & 0.08 ($\downarrow$) & 0.08 ($\downarrow \downarrow$)& 2.00 & 0.99 ($\downarrow$) & 0.47 ($\downarrow \downarrow$)& 42.1\% & 17.6\% ($\downarrow$) & 4.6\% ($\downarrow \downarrow$)\\\cline{2-14}

~ & S-4 & 9 & 9 ($\uparrow$) & 12 ($\uparrow \uparrow$) & 0.12 & 0.08 ($\downarrow$) & 0.04 ($\downarrow \downarrow$)& 2.38 & 1.99 ($\downarrow$) & 0.57 ($\downarrow \downarrow$)& 49.2\% & 36.6\% ($\downarrow$) & 6.4\% ($\downarrow \downarrow$)\\\cline{2-14}

%production ready
\hline
\multirow{17}*{Huawei OceanStor} & S-5 & 582 & 648 ($\uparrow$) & 661 ($\uparrow \uparrow$) & 27.9 & 25.2 ($\downarrow$) & 24.1 ($\downarrow \downarrow$)& 467.2 & 224.1 ($\downarrow$) & 140.2 ($\downarrow \downarrow$)& 31.3\% & 22.5\% ($\downarrow$) & 2.8\% ($\downarrow \downarrow$)\\\cline{2-14}

~ & S-6 & 639 & 764 ($\uparrow$) & 845 ($\uparrow \uparrow$)& 26.8 & 20.9 ($\downarrow$) & 18.9 ($\downarrow \downarrow$)& 276.6 & 191.3 ($\downarrow$) & 150.3 ($\downarrow \downarrow$)&  18.2\% & 6.3\% ($\downarrow$) & 5.8\% ($\downarrow \downarrow$)\\\cline{2-14}

~ & S-7 & 1535 & 1664 ($\uparrow$) & 1662 ($\uparrow \textcolor{red}{\downarrow}$)& 9.9 & 9.6 ($\downarrow$) & 9.6 ($\downarrow \downarrow$)& 284.1 & 163.2 ($\downarrow$) & 154.8 ($\downarrow \downarrow$)& 21.3\% & 4.2\% ($\downarrow$) & 2.3\% ($\downarrow \downarrow$)\\\cline{2-14}

~ & S-8 & 638 & 666 ($\uparrow$) & 731 ($\uparrow \uparrow$)& 219.8 & 195.2 ($\downarrow$) & 176.3 ($\downarrow \downarrow$)& 1380.2 & 956.2 ($\downarrow$) & 561.2 ($\downarrow \downarrow$)&  32.1\% & 25.2\% ($\downarrow$) & 9.9\% ($\downarrow \downarrow$)\\\cline{2-14}

~ & S-9 & 621 & 698 ($\uparrow$) & 966 ($\uparrow \uparrow$)& 189.3 & 185.4 ($\downarrow$) & 132.1 ($\downarrow \downarrow$)& 1872.1 & 1154.3 ($\downarrow$) & 629.2 ($\downarrow \downarrow$)&  38.8\% & 28.6\% ($\downarrow$) & 4.6\% ($\downarrow \downarrow$)\\\cline{2-14}

~ & S-10 & 1322 & 1571 ($\uparrow$) & 1700 ($\uparrow \uparrow$)& 94.5 & 82.3 ($\downarrow$) & 75.9 ($\downarrow \downarrow$)& 1007.8 & 893.7 ($\downarrow$) & 819.8 ($\downarrow \downarrow$) & 20.0\% & 12.6\% ($\downarrow$) & 11.1\% ($\downarrow \downarrow$)\\\cline{2-14}

~ & S-11 & 1272 & 1321 ($\uparrow$) & 1275 ($\uparrow \textcolor{red}{\downarrow}$)& 13.9 & 12.5 ($\downarrow$) & 12.5 ($\downarrow \downarrow$)& 276.3 & 169.3 ($\downarrow$) & 126.7 ($\downarrow \downarrow$)& 19.5\% & 16.8\% ($\downarrow$) & 3.0\% ($\downarrow \downarrow$)\\\cline{2-14}

~ & S-12 & 1566 & 1633 ($\uparrow$) & 1659 ($\uparrow \uparrow$)& 10.7 & 9.9 ($\downarrow$) & 9.6 ($\downarrow \downarrow$)& 199.7 & 166.2 ($\downarrow$) & 134.3 ($\downarrow \downarrow$)&  27.9\% & 13.6\% ($\downarrow$) & 2.3\% ($\downarrow \downarrow$)\\\cline{2-14}

~ & S-13 & 2589 & 2768 ($\uparrow$) & 3006 ($\uparrow \uparrow$)& 6.1 & 5.8 ($\downarrow$) & 5.3 ($\downarrow \downarrow$)& 257.1 & 124.2 ($\downarrow$) & 132.8 ($\downarrow \downarrow$)& 30.2\% & 9.2\% ($\downarrow$) & 2.4\% ($\downarrow \downarrow$)\\\cline{2-14}

~ & S-14 & 1491 & 1539 ($\uparrow$) & 1656 ($\uparrow \uparrow$)& 88.9 & 84.4 ($\downarrow$) & 76.7 ($\downarrow \downarrow$)& 635.4 & 459.9 ($\downarrow$) & 355.6 ($\downarrow \downarrow$)& 23.6\% & 17.9\% ($\downarrow$) & 3.9\% ($\downarrow \downarrow$)\\\cline{2-14}

~ & S-15 & 1889 & 1867 ($\textcolor{red}{\downarrow}$) & 2056 ($\uparrow \uparrow$)& 79.0 & 69.0 ($\downarrow$) & 62.2 ($\downarrow \downarrow$)& 553.0 & 474.5 ($\downarrow$) & 421.8 ($\downarrow \downarrow$)& 18.1\% & 14.7\% ($\downarrow$) & 3.5\% ($\downarrow \downarrow$)\\\cline{2-14}

~ & S-16 & 3170 & 3495 ($\uparrow$) & 3876 ($\uparrow \uparrow$)& 46.1 & 37.4 ($\downarrow$) & 32.8 ($\downarrow \downarrow$)& 633.5 & 533.7 ($\downarrow$) & 478.9 ($\downarrow \downarrow$)& 25.7\% & 14.2\% ($\downarrow$) & 3.3\% ($\downarrow \downarrow$)\\\cline{2-14}

~ & \multicolumn{1}{|c|}{R-A} & 1369 & 1495 ($\uparrow$) & 1696 ($\uparrow \uparrow$)& 10.8 & 10.7 ($\downarrow$) & 9.4 ($\downarrow \downarrow$)& 288.7 & 180.8 ($\downarrow$) & 166.6 ($\downarrow \downarrow$)& 14.9\% & 7.9\% ($\downarrow$) & 3.0\% ($\downarrow \downarrow$)\\\cline{2-14}

~ & \multicolumn{1}{|c|}{R-B} & 1493 & 1573 ($\uparrow$) & 1841 ($\uparrow \uparrow$)& 93.2 & 82.1 ($\downarrow$) & 69.3 ($\downarrow \downarrow$)& 1590.2 & 1148.4 ($\downarrow$) & 840.2 ($\downarrow \downarrow$)& 19.6\% & 11.5\% ($\downarrow$) & 3.9\% ($\downarrow \downarrow$)\\\cline{2-14}

~ & \multicolumn{1}{|c|}{R-C} & 392 & 424 ($\uparrow$) & 479 ($\uparrow \uparrow$)& 9.8 & 9.4 ($\downarrow$) & 8.3 ($\downarrow \downarrow$)& 196.9 & 107.7 ($\downarrow$) & 104.7 ($\downarrow \downarrow$)& 24.5\% & 4.8\% ($\downarrow$) & 3.4\% ($\downarrow \downarrow$)\\\cline{2-14}

~ & \multicolumn{1}{|c|}{R-D} & 306 & 310 ($\uparrow$) & 329 ($\uparrow \uparrow$)& 1.1 & 0.8 ($\downarrow$) & 0.7 ($\downarrow \downarrow$)& 123.1 & 65.6 ($\downarrow$) & 48.5 ($\downarrow \downarrow$)& 22.9\% & 7.4\% ($\downarrow$) & 3.6\% ($\downarrow \downarrow$)\\\cline{2-14}

~ & \multicolumn{1}{|c|}{R-E} & 489 & 562 ($\uparrow$) & 670 ($\uparrow \uparrow$)& 7.6 & 7.1 ($\downarrow$) & 6.0 ($\downarrow \downarrow$)&  276.6 & 125.8 ($\downarrow$) & 114.8 ($\downarrow \downarrow$)& 36.7\% & 15.7\% ($\downarrow$) & 4.6\% ($\downarrow \downarrow$)\\
\hline
\end{tabular}
}
\vspace{-0.6em}
\end{table*}

\vspace{-1.2em}
\subsection{Overall Evaluation}
%We first show the performance comparisons among learning based framework, rule-based methods and the system without control method. Then %the same performance comparison was conducted using five real-world workloads. Lastly
%, we evaluate the efficient of \LQoCo.

\subsubsection{Evaluation on different workloads}
% Then, we analyze and discuss the learning capability of \LQoCo under different workloads and meanwhile evaluate the advantage of the domain knowledge and adaptive bound mechanism we designed for \LQoCo. Lastly, we evaluate the adaptability of \LQoCo.

% \subsubsection{Effectiveness comparison}
\label{Overall}

%On OceanStor, we evaluated our \LQoCo with \CoTo and \NoCoTo. Besides, \LQoCo was compared with `bypass' and `No-bypass' on Linux platform. %All above comparisons were performed using those workloads mentioned in Section~\ref{meth}. %By means of Vdbench, We set different block size (8K and 32K), I/O Size (32K and 256K) and R/W Ratio (0\%, 30\% and 70\%) for generating different synthetic I/O workloads. Besides, five workloads collected from the real world are also tested with above control mechanisms.
%Note that to ensure the above workloads can overload the cache capacity in the storage system, we employ enough concurrent threads for each workload.
The detailed results are recorded in Table~\ref{overallresult}. We initialize the Q table (i.e., training the model from the scratch) for each testing workload. It can be seen that our designed learned framework \LQoCo achieves higher throughput and lower throughput jitter than the systems without control methods in all cases and outperforms these rule-based methods in almost all cases. %To be more intuitive, four cases (two from Linux and two from OceanStor) from Table~\ref{overallresult} are visualized in Figure~\ref{fig:fourcases}, where 
%Besides, \LQoCo outperforms both \CoTo and `bypass' and the system without cache capacity overload control methods.
% Like the generality of our \LQoCo, we also applied it to the distributed storage systems and got some expected preliminary results, which are not described here due to space limitations.
This is because that \LQoCo uses the trial-and-error manner in RL and employs exploration \& exploitation strategies to explore more optimal bandwidth, thus making full use of the storage tier's storage ability while effectively preventing cache overloading. 
%Also, the token bucket scheme shapes the random I/O requests and makes them properly confined to the storage cache between the storage backend and the host-side resulting in the low throughput jitter. 
Also, as the improper bandwidth employed by \CoTo ('bypass') and \NoCoTo ('No-bypass') for overloading control, more resources (e.g., CPU) are used for I/O queue processing.
As a result, fewer resources are used for I/O requests execution, which leads to throughput degradation~\cite{Inho2020}. Besides, \LQoCo also achieves lower mean latency and especially gains a huge improvement on the tail latency. Tail latency is largely affected by queuing, and overloading has huge negative impacts for all workloads sharing the underlying infrastructure~\cite{xie2018cutting}. In summary, we could claim that \LQoCo can greatly optimize the performance of storage systems.

\subsubsection{Efficiency comparison}

\iffalse
\begin{table}[t]\vspace{-1em}
\begin{center}
\caption{The cost comparison for \LQoCo, \CoTo (bypass) and \NoCoTo (No-bypass). \LQoCo cost more but acceptable. }
\scalebox{0.75}
{
\begin{tabular}{|c|c|c|c|c|}
\hline
Platform & Method & Training & Memory Cost & Computing Cost (in 1 sec.) \\
\hline
\multirow{3}*{Linux} &\LQoCo & 10 mins & 100KB & 12 us \\
\cline{2-5}
~ & bypass & 0 & 0 & 1 us \\
\cline{2-5}
~ & No-bypass & 0 & 0 & 0 \\
\hline
\multirow{3}*{OceanStor} &\LQoCo & 10 mins & 100KB & 100 us \\
\cline{2-5}
~ & \CoTo & 0 & 0 & 50 us \\
\cline{2-5}
~ & \NoCoTo & 0 & 0 & 0 \\
\hline
\end{tabular}
}
\label{cost}
\end{center}
\end{table}
\fi

%In this section, we evaluate the efficiency of \LQoCo, from the perspective of the model training time, memory cost and computing cost, compared with \CoTo and \NoCoTo. Table~\ref{cost} provides further statistical results for each method.

For training time, \LQoCo takes around 10 minutes (the time from the model initialization to the time when the system performance tends to be stable) for model training. %(from initialization of model to the convergence of model training) 
%whereas \CoTo and \NoCoTo do not need. 
Note that \LQoCo only needs training once and will subsequently be used and updated to recommend bandwidth %for cache capacity overloading control 
in storage systems. Such training time for a RL model is acceptable in storage systems.
For memory cost, we %As we described in Section~\ref{subsub:training}, the Q table is the policy in RL which drives our \LQoCo to work. 
need to store and update the Q table in memory. %The size of the Q table is determined by the number of the rows (states) and columns (actions) we designed in \LQoCo. 
The Q table used in \LQoCo only occupies about 100 KB memory which is very light for storage systems.
For computing cost, %The computing cost consists of system state collection, online model updating and the execution time in the token bucket. 
we analyze the time \CoTo and \LQoCo spend in each second. \LQoCo takes about 100 us and \CoTo takes about 50 us which are almost negligible compare to other computing costs. From these three aspects, it can be seen \LQoCo is a light-weight cache management technique suitable for storage systems. %Note that the training time cost can be further shortened if model compression and acceleration technologies~\cite{choudhary2020comprehensive} are used. We leave it for our future work to explore other better solutions for further optimizing the efficiency of our proposed method.

%\vspace{-1.2em}
%\subsection{Evaluation for more intuitions}
\label{intuitively}

\begin{figure}
\centering
%\subfigure[Comparison between systems with/without cache overload method in Linux. ]{
%\includegraphics[width=3.2in]{./fig/preliminary_linux.pdf}
%}
%\quad
%\subfigure[Comparison between systems with/without cache overload method in OceanStor.]
{
\includegraphics[width=3.2in]{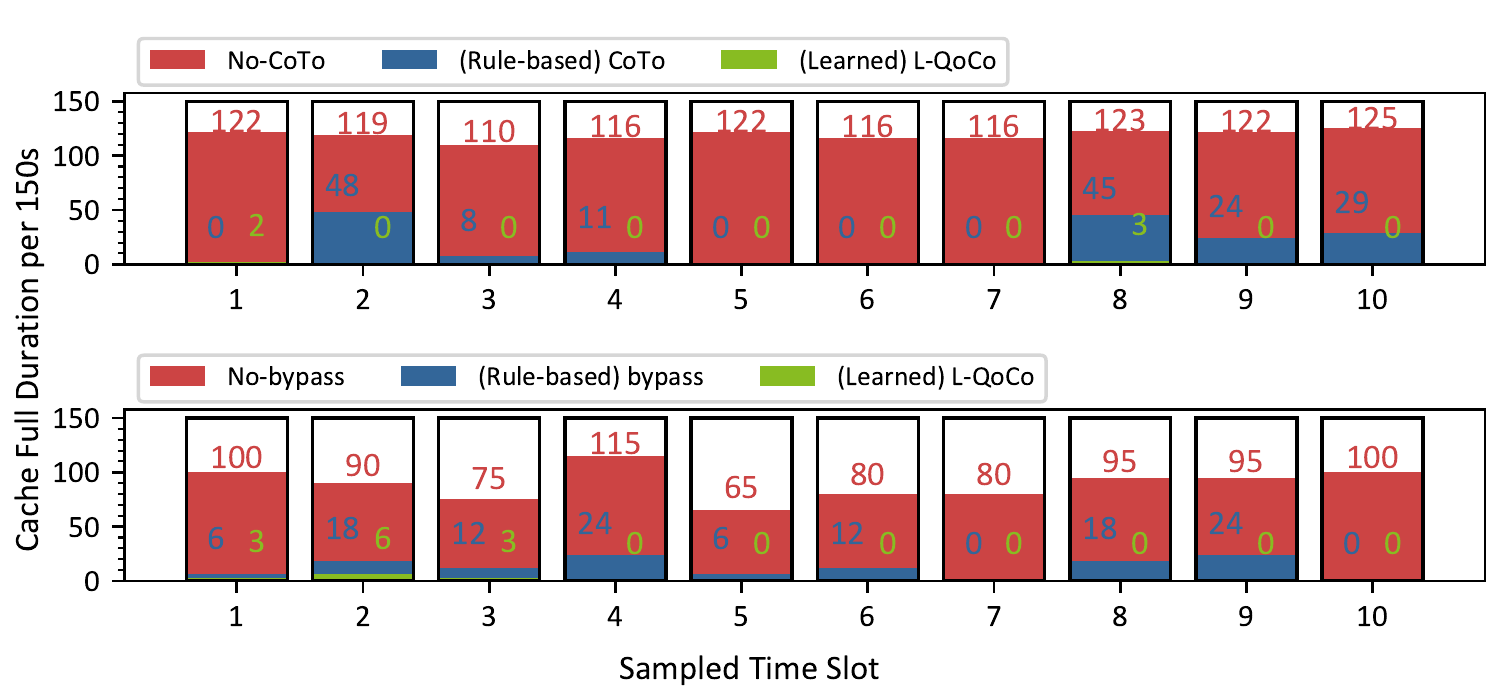}
}
\caption{The number of cache full duration in each 150 secs.}
\label{fig:full}
\end{figure}

\subsubsection{Cache full duration} We investigate the duration of cache being full in each 150s. The visualized results are recorded in Figure~\ref{fig:full}. To align with the 150s time window, for each workload, the duration of 1500s is split up into 10 equal-length time slots. The horizontal axis presents 10 sampled time slots and the vertical axis represents the duration of the cache being full with/without the cache overloading control method. It can be seen that the cache with \LQoCo is with much shorter duration of the cache being full, compared to the two rule-based methods and one without cache overloading control. Thus, \LQoCo can effectively control the bandwidth to prevent cache overloading. 

% \vspace{-1em}
\subsection{Adaptability Comparison}
\begin{figure}[t]%\vspace{-0.2em}
\centering
\includegraphics[width=3.35in]{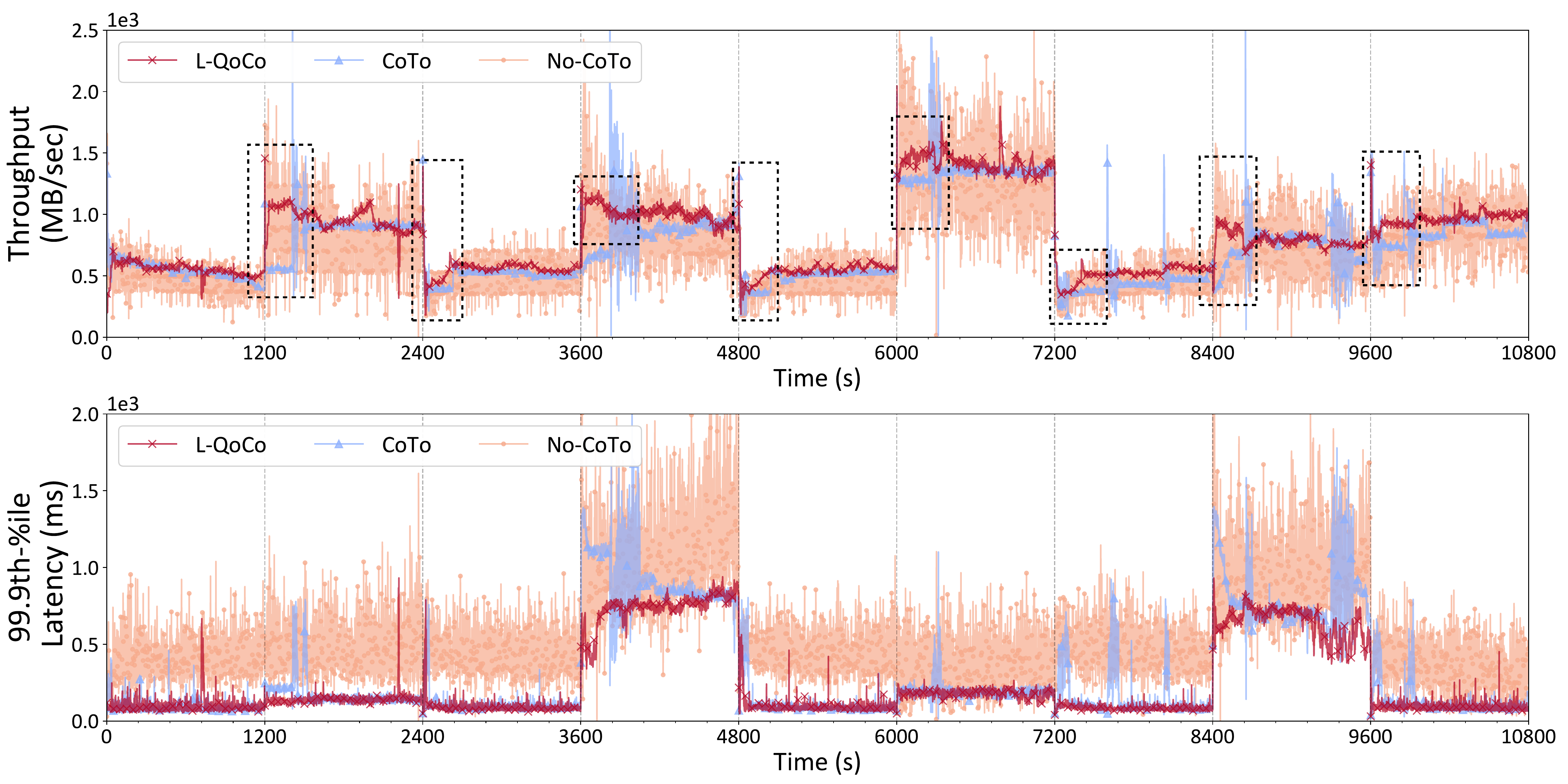}
\caption{Performance comparisons among learned framework and the system with/without rule-based method. }
\label{fig:complex}
\end{figure}

To verify the adaptability of \LQoCo,
we compare it with \CoTo and \NoCoTo under the constantly-changing workloads. The order in which we replay the workloads is: S-17$\rightarrow$S-6$\rightarrow$S-18$\rightarrow$S-9$\rightarrow$S-18$\rightarrow$S-7$\rightarrow$S-18$\rightarrow$S-8$\rightarrow$S-19. It should be noted that at least one workload property (Block Size, I/O Size and R/W Ratio) changes between each changing workload. %Besides, note that we only initiate the Q table once at the beginning of the workloads, which is different from the experiments we employed in Section~\ref{Overall}.
Figure~\ref{fig:complex} shows that for these complex workloads, the performance of \LQoCo exhibits higher throughput, lower throughput jitter and tail latency compared to \CoTo and \NoCoTo in all workloads. Besides, \LQoCo solves the problem that the tail latency becomes large suddenly if \CoTo recommends an improper bandwidth (this problem as shown in Figure~\ref{fig:withandwithout}).
%It is worthy to find that \LQoCo only performs several large 99.9th-\%ile latency early in the experiment because the RL model in \LQoCo is just starting to learn and this situation disappeared quickly after the model converged. 
The recommendation policy of \CoTo is adjusted in step-by-step mode but \LQoCo can recommend a proper bandwidth faster than \CoTo during the workload changing for improving system performance and stability. %We marked the changing position in Figure~\ref{fig:complex} for further observation.
We attribute this to the fact that \LQoCo applied the RL so it can  discover the better bandwidth more accurately and adaptively by either exploiting current knowledge or exploring unknown states to maximize a cumulative reward, thus reducing the possibility of falling into a local optimum. %The experimental results manifest that our designed \LQoCo is with strong adaptability to time-varying workload without retraining a new model that learns towards an optimizing direction and recommends reasonable bandwidth limiting corresponding to the current system states for different workloads.

\vspace{-1em}
\subsection{Acceleration and Robustness Improvement}
\label{KnowledgeandSafeguard}

\begin{figure}[t]%\vspace{-0.36em}
\centering
\includegraphics[width=3.35in]{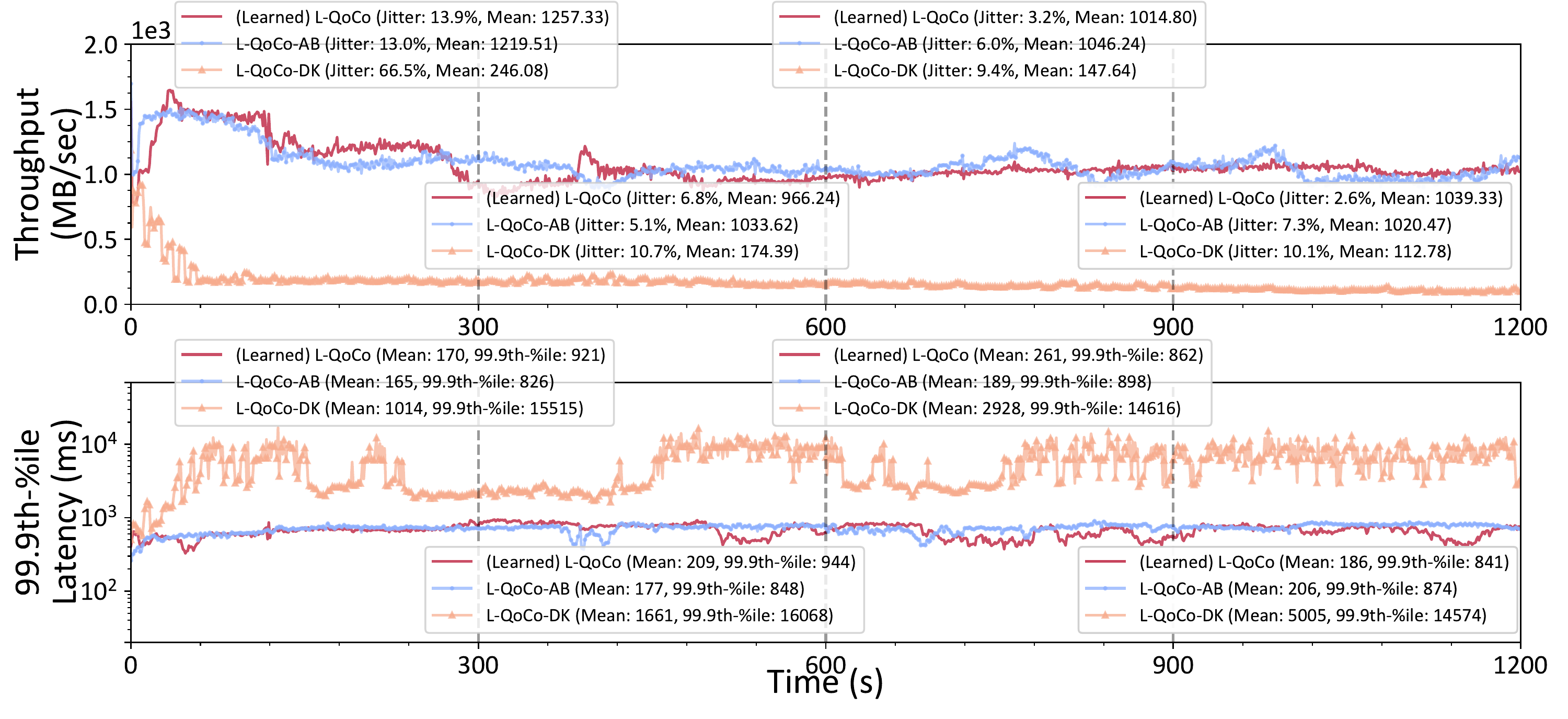}
\caption{Performance comparison for \LQoCo, \DK and \SB to verify the effectiveness of our proposed mechanism. }
\label{fig:ability}
\end{figure}

To verify the benefits of acceleration of learning using domain knowledge and adaptive bound mechanism, a comprehensive ablation evaluation has been performed.
We compare \LQoCo with its two variants, \LQoCo without domain knowledge (\DK) and \LQoCo without adaptive bound (\SB). As shown in Figure~\ref{fig:ability}, \LQoCo achieves stable and higher performance quickly.
\SB (without adaptive bound), although it performs similar results with \LQoCo, larger throughput jitters and P99.0 latency can be observed, which indicates that it is hard for model convergence.
This is because \SB performs uncontrollable exploration (we described in Section~\ref{Safeguard}) resulting in storage system performance fluctuation.
%We marked them out for highlight in Figure~\ref{fig:ability}. 
For \DK (without domain knowledge), we find that it performs the worst performance.
This demonstrates that \DK makes the model hard for convergence and cannot recommend the proper overall bandwidth adaptively for the storage system. %Besides, we admit that the \DK applied the RL should have the ability to learn even without domain knowledge, but we cannot find this phenomenon in this figure.
If we look at the throughput curve of \DK in-depth, we can find that this situation is caused by our designed actions in RL that using the fixed percentage for an adjustment (i.e., FastDecrease is -3\% and FastIncrease is 3\% which we described in Section~\ref{RL4L}). In detail, it can be observed that \DK is eager to get close to the proper bandwidth and therefore makes a lot of actions like increasing at the first 60 seconds.
However, an increasing action may still put the storage system in worse states (cache watermark is low) and get a negative reward resulting in alternating actions of increasing and decreasing. As the same percentage for adjustment, the overall trend of the bandwidth recommended by \DK is decreasing and cannot escape from this situation. On the contrary, this is solved by \LQoCo by including domain knowledge. %Considering the simplicity of the fixed percentage for action adjustment and the effectiveness after we add the domain knowledge for better model training, we finally adopt this method in our \LQoCo. %Note that it is interesting to investigate other flexible (not fixed) adjustment strategies in \LQoCo for achieving better performance. 
In summary,by including domain knowledge and the adaptive bound mechanism into \LQoCo, we can effectively speedup model training and further improve the storage system performance.

\vspace{-0.6em}
\section{Conclusion}
\label{sec:C}
In this paper, we first formally define the cache overloading in storage systems and %design a cache-oriented capacity overload control mechanism \CoTo in our designed storage system (commonly used in practical) on Huawei OceanStor, which uses the cache watermark level to form a finite state machine for capacity overload control for the cache. The results from preliminary studies show that \CoTo greatly improves the performance compared to the system without overload control mechanism. However, the rule-based control mechanisms (both \CoTo and `bypass') are not adaptive, which might still degrade the performance of the storage system. Thus, we 
propose an online RL based cache bandwidth control technique, called \LQoCo. We design an effective reward function and incorporate useful domain knowledge into \LQoCo which greatly accelerates the model learning of RL. Besides, an adaptive bound strategy is designed to further improve the robustness of \LQoCo. Extensive experiments conducted in the Linux platform and the Huawei OceanStor storage system show that \LQoCo outperforms the existing rule-based techniques. 

%\clearpage
% performed superior bandwidth control for various workloads, which greatly improved the storage system performance. We also demonstrated \LQoCo had good adaptability whenever the workloads changed. 

%-------------------------------------------------------------------------------

%\bibliographystyle{ACM-Reference-Format}
\bibliographystyle{unsrt}
\vspace{-0.6em}
\bibliography{reference}
\end{document}